\documentclass[twocolumn,amsmath,amssymb]{revtex4-1}
\usepackage[dvips]{graphicx}
\setkeys{Gin}{width=8.5cm}

\begin{document}
\title{Self-shaping dynamical systems and learning}
\author{Natalia~B.~Janson}
\email[E-mail: ]{N.B.Janson@lboro.ac.uk}
\author{Christopher~J.~Marsden}
\affiliation{School of Mathematics, Loughborough University, Loughborough LE11 3TU, UK}

\begin{abstract} 

We associate learning and adaptation in living systems with the shaping of the velocity vector field in the respective dynamical systems in response to external, generally random, stimuli. With this, a  mathematical concept of self-shaping dynamical systems is proposed.  Initially there is a zero vector field and an ``empty" phase space with no attractors or other non-trivial objects. As the random stimulus begins, the vector field deforms and eventually becomes 
smooth and deterministic, despite the random nature of the applied force, while the phase space develops various geometrical objects. We consider gradient self-shaping systems, whose vector field is the gradient of some energy function, which under certain conditions develops into the multi-dimensional probability density distribution (PDD) of the input. Self-shaping systems are relevant to neural networks (NNs) of two types: Hopfield, and probabilistic. Firstly, we show that they can potentially perform pattern recognition tasks traditionally delegated to Hopfield NNs, but without supervision and on-line, and without developing spurious minima of the energy. Secondly, like probabilistic NNs, they can reconstruct the PDD of input signals, without the limitation that new training patterns have to enter as new hardware units. Thus, self-shaping systems can be regarded as a generalization of the NN concept, achieved by abandoning the ``rigid units" - ``flexible couplings" paradigm and making the vector field fully flexible and amenable to external force. The new concept presents an engineering challenge requiring new principles of hardware design. It might also become an alternative paradigm for modeling of living and learning systems. 

\end{abstract}
 
\maketitle

\section{Introduction}
In the past century there occurred a revolution in terms of mathematical understanding of biological systems: their {\it dynamical} nature was appreciated at all levels of organization, from single cells, through organisms, and to the populations of organisms, meaning that their state is not static, but is continuously {\it changing} in time. In particular,  the generality and persistence of oscillations in living systems has been widely acknowledged. Just a few examples include pacemaker cells and neuron firings at the cellular level, 
heart beats, breathing and circadian rhythms at the level of an organism, and fluctuations in population size in the communities of organisms.  Since then, living systems have  often been modelled as dynamical systems.  A concept of a dynamical disease was proposed \cite{Dyn_disease1}, and nowadays  new medicines require testing with mathematical models before their mass-production is approved \cite{Testing_medicines}. 

A dynamical system is a mathematical construction incorporating  a vector $x=(x_1,\ldots,x_N)$, that describes the system state at any time moment $t$,  and some rule that determines how the state evolves in time. This evolution rule can be defined, e.g. by a system of ordinary differential equations, 
\begin{equation}
\label{ds}
\frac{\mathrm{d} x_1}{\mathrm{d} t}=s_1(x_1,\ldots,x_N),  \ \ \ldots, \ \  
\frac{\mathrm{d} x_N}{\mathrm{d} t}=s_N(x_1,\ldots,x_N). 
\end{equation}
Here, $s=(s_1,\ldots,s_N)$ is a {\it phase velocity vector field}, 
which can be loosely understood as a  ``force" that pushes the state $x(t)$ in a certain direction, and is generally different at different positions in the state space. Remarkably, even if the vector field $s$ is permanently fixed at all points, it generally makes the state change, i.e. creates the ``behavior". 

Crucially, all living systems are {\it dissipative} because they permanently lose energy as they function. Mathematically, they can be described by dissipative dynamical systems that have attractors: geometrical objects in the phase space to which all solutions converge from a certain vicinity \cite{MacKey_Glass_clocks}. Attractors are very important in the context of self-organization: a dissipative system can be launched from a randomly chosen initial condition, but with time its behavior will automatically settle down on the same stationary mode, whose geometrical image is an attractor. 

The most prominent feature of all living systems is their ability to 
{\it modify} themselves under the influence of the environment. An extreme example would be a lizard that grows a new tail after the old one is lost. Some more common  examples include the growth of frequently used muscles, the development of  stamina in response to exercise, and increasing the flexibility of the joints in response to their stretching. Importantly, the environmental influence is generally quite {\it random}, but the living system responds to it in a coherent manner. 
With account of this adaptation ability, it might be more appropriate to model living systems as dynamical systems, whose vector field {\it modifies} itself in time automatically in response to the external random stimulus. 

{\bf Learning in the brain.} The most striking feature of a sufficiently advanced living system is its ability to {\it learn}. Learning mechanisms in living systems are associated with the nervous system: the brain and its connections with all parts of the body. Since the first discovery that the brain does not represent a homogeneous substance, but is rather a collection of intertwined discrete units called neurons \cite{Neuron_doctrine},  a huge volume of biological and psychological research has been carried out in order to reveal the biological mechanisms of learning. It is well established that in the course of learning the {\it architecture} of the brain changes.  Namely, while the internal structure of the individual neurons remains roughly the same, the {\it connections} between different neurons change in time in response both to the sensor stimuli, and to the processes inside the brain \cite{Synaptic_plasticity}. This fact has given rise to a separate research area in the field of artificial intelligence: artificial neural networks. At the same time, it contributed to the cognitive theory, and to the philosophy of science in general, by giving birth to the connectionism paradigm \cite{Connectionism}, within which all knowledge (or {\it information}) in the brain is represented in the form of the strengths of connections between the neurons. Note, that the sensory stimuli that the brain receives are typically quite {\it random}, but the brain seems to accumulate information in a consistent and orderly manner. 

{\bf Information.} We point out that while the term ``information" has penetrated all spheres of human activity and is used most broadly, we are still lacking an accurate and at the same time sufficiently broad definition of it. Information theory, which has been introduced and developed within mathematical and physical sciences, operates with sequences of symbols and various probabilities of their occurrence.  There are a few definitions of information, and the most widely used seem to be those proposed by Shannon \cite{Shannon_info} and Fisher \cite{Fisher_info}. Where a message cannot be reduced to a sequence of symbols, there is no suitable mathematical theory. 

One example illustrating the limitations of modern information theory is our perception of facial expression, e.g. a smile. 
While it might be easy to classify the message as a smile, the subtle {\it meaning} of it might vary considerably, from approving to ridiculing. 
An ideal information theory should be able to detect all the {\it meanings} in the message together with their relative quantities. 
Another general problem of scientific and philosophical thought is the relationship between information, energy and matter  \cite{Szilard_info_energy,Info_matter}. 

Within this paper we do not aim to contribute to the proper development of a meaning-based information theory, or to resolve the  debate above. However, we propose a somewhat broader definition of information, which we feel could be useful for the practical purposes of this paper, and would contribute to the ``matter--information" debate. 

Consider a simple example: a sequence of symbols can be written on paper, on the sand, or made of concrete blocks. Regardless of the material used, the message contains exactly the same amount of information. 
Therefore, it is the {\it shape} that the material object takes, that can be called information. 
The shape can be certainly understood quite {\it broadly}, not only as a geometrical shape of a material object, but also as its architecture or internal structure. E.g. the shape of an envelope of high-frequency electromagnetic waves  can carry the same information as the sound perceived as mechanical oscillations of an ear membrane. 

{\bf Definition.} Information is the {\it shape of the matter}. 

{\bf Learning and shaping.} If learning can be understood as acquiring information, for practical (e.g. engineering) purposes we define learning as {\it changing the shape} of the system in response to the external stimulus. 

{\bf Learning by a dynamical system.} For the rest of the paper we will stay within the framework of {\it dynamical systems theory}. 
{\bf Definition.}  {\it Learning} by a dynamical system is the shaping of its {\it velocity vector field} in response to external stimuli and/or internal processes. 

{\bf Goal.} We wish to construct a dynamical system (\ref{ds}) experiencing a continuous, generally random, external force, and allow this force to systematically {\it deform} the velocity vector field according to a certain rule. The external influence should accumulate and, despite its random nature, give rise to a {\it smooth} vector field, which could eventually become fully {\it deterministic} and highly organized, and thus give rise to a new behavior of the dynamical system. Importantly, the resulting structure of the vector flow in the system should be determined by the {\it statistical} properties of the random input. We propose to call such systems {\it self-shaping dynamical systems}. 

Self-shaping systems would be different from the well-known random dynamical systems of the form $\frac{\mathrm{d} x}{\mathrm{d} t}=q(x,\xi(t))$, in which $\xi(t)$ is a random input and $q(x,0)=s(x)$ with $s(x)$ being the vector field from (\ref{ds}) \cite{Freidlin_random_pert}. In the latter systems the random input only {\it perturbs} the existing vector field, while in the self-shaping systems the vector field will be {\it created} by the random input. 

\section{Gradient self-shaping systems}

In this paper we concentrate on the simplest form of the self-shaping systems, the so-called gradient (or potential) systems, in which the vector field $s$ is the gradient of a certain energy function $V$, 
\begin{equation}
\label{eq_particle}
\frac{\mathrm{d} x}{\mathrm{d} t}=-\frac{\partial V(x,t)}{\partial x},
\end{equation}
where $x$ represents the location in $N$-dimensional space.  The state point in such a system behaves just like a massless particle that is placed into a potential energy landscape $V(x)$, which moves towards the relevant local minimum. Here, we assume that the energy $V$ is also a function of time $t$, 
to take into account the continuous shaping process. 

\begin{figure}
\includegraphics[width=0.4\textwidth]{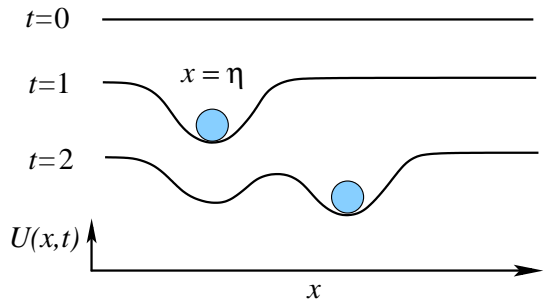}
\caption{\label{fig_foam_ill} Illustration of the idea of the flexible energy landscape as a memory foam. 
For a one-dimensional  ``foam" stretched in the $x$ direction, assume that initially it is flat, i.e. its landscape is described as $U(x,0)$$=$$0$ (see $t$$=$$0$). If a stone drops onto the foam at position $x$$=$$\eta$, the  
landscape is deformed: a dent appears, which is the deepest exactly at 
$x$$=$$ \eta$, and gets shallower at larger distances from $\eta$ (see $t$$=$$1$). In other words, the foam will learn about the occurrence of the stone and of its position. 
}
\end{figure}

Below we derive an equation describing the shaping of the energy $V$ in response to the random stimulus. It is helpful to employ 
a loose analogy with the ``memory foam" used in orthopedic mattresses. This foam takes the shape of a body pressed against it, but slowly returns to its original shape after the pressure is removed.  It helps to use the auxiliary function $U(x,t)$ describing the foam landscape, as illustrated by Fig. \ref{fig_foam_ill}. 
Also, assume that the foam is elastic with elasticity factor $k$ that models the capacity of the system to forget. Here, we make a simplified assumption that the deeper the dent at the position $x$ is, the faster the foam tries to come back to $U$$=$$0$. However, the forgetting term can be modelled in a variety of ways, depending on what the situation requires. 

Now assume that we subject the foam to a continually varying external stimulus $\eta (t)$, as if at any new time moment $t$ a new stone drops at a new position $x$$=$$\eta(t)$ (Fig. \ref{fig_foam_ill}, $t$$=$$2$). Thus the ``foam" will undergo a continuous shaping process.  The signal  $\eta (t)$ can be of either deterministic, or stochastic nature, and can have arbitrary statistical properties.

Consider how the foam landscape changes over a small, but finite time interval $\Delta t$:
\begin{equation}
\label{eq1}
U (x, t +\Delta t) = U (x, t) - g(x - \eta (t)) \Delta t - k U (x, t) \Delta t, 
   \end{equation}
where $g(z)$ is some non-negative bell-shaped function, describing the shape of a single dent, e.g. a Gaussian function,
\begin{equation} 
g(z)=\frac{1}{\sqrt{2 \pi \sigma^2_z}} \exp \bigg( - \frac{z^2}{\sigma^2_z} \bigg).
\end{equation}


In (\ref{eq1})  move $U (x, t)$ to the left-hand side, divide both parts by $\Delta t$, and take the
limit as $\Delta t \rightarrow 0$, to obtain 
\begin{equation}
\label{eq2}
  \frac{\partial U (x, t)}{\partial t} = - g(x - \eta (t)) - k U (x, t).
\end{equation}
It can be shown by numerical simulation with some arbitrary $\eta (t)$, that the solution $U (x, t)$ has
a linear trend, i.e. it behaves as a linearly decaying function of $t$ with superimposed 
fluctuations. We wish to eliminate this trend and see if we can achieve some sort of stationary behavior of $U (x, t)$. Perform the change of variables
\begin{eqnarray*}
V = \frac{U}{t}, \quad \frac{\partial V}{\partial t} =  \frac{1}{t} \left(
   \frac{\partial U}{\partial t} - V \right), \quad 
  \frac{\partial U}{\partial t} = t \frac{\partial V}{\partial t} + V, 
\end{eqnarray*}
and rewrite (\ref{eq2}) as follows
\begin{equation}
\label{eq3}
  \frac{\partial V}{\partial t} = - \frac{1}{t} \bigg( V + g(x - \eta)  \bigg) - k V.
\end{equation}
Within this model, the energy landscape and the vector field of Eq. (\ref{eq_particle}) progressively smooth out and stabilize, as illustrated in Fig. 2, 
if $\eta(t)$ is a stationary and ergodic process. 


\noindent {\bf Proof of shaping into the input density.}  Next, we prove that under certain conditions listed below, the energy landscape $V$ of (\ref{eq_particle}) automatically shapes into the negative of the probability density distribution of the input random process. 

Consider the evolution of $V(x,t)$, where the $N$-dimensional input vector $\eta(t)$ is a realization of a strict-sense {\it stationary} and {\it ergodic} random process $H(t)$ with some arbitrary probability density distribution (PDD) $p_N^{H}(\eta_1, \eta_2, \ldots, \eta_N)$. Due to {\it stationarity}, $p_N^{H}$ does not change in time; due to {\it ergodicity}, any single realization $\eta(t)$ contains all information about $p_N^{H}$, i.e. any statistical characteristic can be obtained from $\eta(t)$ by averaging over time, rather than over the ensemble of realizations that would have been required for a non-ergodic process \cite{Stratonovich_vol1}. Below we will show that with time, $V$ takes the shape of $p_N^{H}$. 

Assume that $k = 0$, i.e. that the system (\ref{eq3}) does not forget what it learnt. 
Multiply both parts of Eq. (\ref{eq3}) by
$\mathrm{d} t$ and integrate. A stationary behavior of $V$ implies 
\begin{equation}
\label{eq4}
\frac{\partial V}{\partial t} = 0, \quad \textrm{and therefore} \quad 
  \int^{\infty}_{-
  \infty} \frac{\partial V}{\partial t} \mathrm{d} t = 0.
\end{equation}
Consider the  integral of the right-hand side of Eq.~(\ref{eq3}) and its limit as $t \rightarrow \infty$
\begin{equation} 
\label{eq_time}
\lim_{t \rightarrow \infty}  \bigg(  - \frac{1}{t} \int^{\infty}_{- \infty}   \big( V + g (x - \eta) \big) \mathrm{d} t \bigg)  \end{equation}
representing the (negative of the) time average $\langle  V + g (x - \eta) \rangle$ of 
the expression under the
integral. The term $g (x - H)$ is a non-linear smooth function of an ergodic process $H$. As proved in \cite{Wolf_ergodic_67}, ``zero-memory nonlinear operations on ergodic processes are ergodic" -- therefore, $g (x - H)$ is also an ergodic random process.  
Thus we can replace the time average (\ref{eq_time}) by the statistical average,
\begin{equation}
\label{eq5}
 \overline{\left( V + g(x - H) \right)} = 
 \int^{\infty}_{- \infty} V p^{H}_{N} (\eta) d \eta + \int^{\infty}_{-
 \infty} g(x - \eta) p^{H}_{N} (\eta) d \eta . 
\end{equation}
In the above, the integral with respect to $\eta$ represents, for brevity, $N$ integrals with respect to the components $\eta_1, \ldots,\eta_N$ of vector $\eta$. 
Since $V$ does not depend on $\eta$ explicitly, the first
term in the right-hand side of (\ref{eq5}) is equal to $V$. The second term is the convolution of  $p^{H}_{N}
(\eta)$ with the function $g(\eta)$. If $g(x - \eta) = \delta (x -
\eta)$, where $\delta(z)$ is Dirac delta-function of several variables, this term is equal to minus
$p^{H}_{N} (x)$, due to the sifting property of delta-function \cite{Bracewell_86}. 
From (\ref{eq3}) combined with (\ref{eq4}) it follows that the expression (\ref{eq5}) is equal to $0$. We therefore
proved that as time $t$ goes to infinity, 
$V(x, t)$ tends to $- p^{H}_{N} (x)$, 
provided that $g(z)$ tends to the Dirac delta-function.  

\begin{figure}
\includegraphics[width=0.45\textwidth]{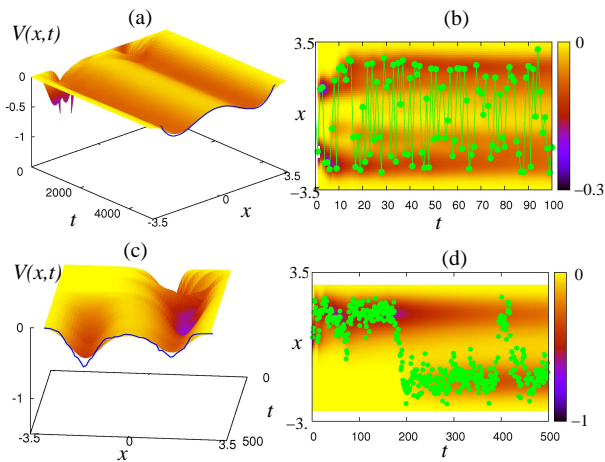}
\caption{\label{fig_uncor_cor} Evolution of the energy landscape $V(x,t)$ as the random stimulus is applied by numerically simulating Eq.~(\ref{eq3}): (a,c) 3D view; (b,d) projection of $V(x,t)$ onto $(x,t)$ plane shown by color (shade of grey), and the stimulus applied -- by filled circles. In (a,c) the probability density distribution of stimulus is given by solid line at the front. 
In (a,b) the consecutive values of the stimulus are uncorrelated, and in (c,d) -- correlated. }
\end{figure}

\noindent {\bf Illustration of shaping into the input  density.} In Fig. \ref{fig_uncor_cor} the evolution of $V(x,t)$ is illustrated, as two kinds of scalar stimuli are applied to the one-dimensional  system (\ref{eq3}). Their PDDs are of similar two-peak shape (see solid lines at the front in (a,c)), but  two consecutive values are non-correlated in (a,b), and correlated in (c,d). 
The stimulus illustrated in Fig. \ref{fig_uncor_cor} (a,b) is obtained by taking Gaussian white noise and applying a non-linear transformation, that   changed its PDD. Thus, the PDD took the shape shown in (a) by solid line, but the consecutive values remained   uncorrelated. The stimulus in (c,d) is obtained by applying Gaussian white  noise to a differential equation describing a particle moving in a   non-symmetric double-well potential with large viscosity \cite{Malakhov_97}. The PDD of the output signal has the shape shown in (c) by solid line, and the consecutive values are correlated.

The actual signals applied are shown by filled circles in (b,d), and in $g(z)$ we used $\sigma_z$$=$$\sqrt{0.1}$. One can see that eventually both energies shape into the respective PDDs, but if the stimulus values are uncorrelated, the convergence is faster. 

If the random process $H(t)$ is not stationary, the energy $V$ evolves into a {\it time-averaged} density of the input. 

\noindent {\bf Relevance to kernel density estimation.} The shaping mechanism which we employed for gradient systems is related to the kernel density estimation used in statistics  \cite{Scott_kernel_92}. Here, we incorporated this mechanism into the continuous dynamical shaping of the vector field, which is done for the first time to the best of our knowledge. Also, the standard assumptions about the kernel density estimators include the statistical {\it independence} of the successive values of the input. Namely, a sequence of input numbers/vectors is regarded as a collection of the values of some random (scalar or vector) {\it variable} with a certain PDD. The convergence to this PDD was proved under these simplifying assumptions only. Here, we prove the convergence to the PDD under a more general assumption, that the successive input values are generated by a random {\it process} and can be correlated with each other. The only requirements used are those of stationarity and ergodicity of this process. 

\section{Relevance to the neural networks}

The  self-shaping systems are in a sense an extension  of a neural network (NN) paradigm. 
In spite of the steadily growing volume of neuroscience research, it would be too premature to claim that we can confidently explain how exactly biological 
NNs function. However, the most essential features of biological NNs seem to be captured by {\it artificial} NNs and their mathematical models.  
Firstly, either biological or artificial NNs are made up of a large number of units (neurons), each with a fixed structure. Notably, it is assumed that one cannot amend the inner structure of individual units. Secondly, these neurons are coupled together through the synaptic connections. Unlike the individual neurons, the couplings can change in the course of time. Namely, new connections can be formed, the old ones can disappear, and the strengths of all connections can change either spontaneously, like in biological NNs, or by a certain pre-defined algorithm, like in artificial NNs. 
This ability is called synaptic plasticity and is associated with the ability to learn. 

Below we demonstrate how self-shaping dynamical systems are related to the two types of NNs: Hopfield and probabilistic ones. 

\subsection{Hopfield neural networks}

~\ Consider a collection of 
one-dimensional ``neurons", whose states can be any real numbers.  An example would be a Hopfield continuous-time NN that can be written, e.g. as follows \cite{Hopfield_NN_85,Hopfield_attractors_86}:
\begin{eqnarray}
\label{NN}
\frac{\mathrm{d} x_i}{\mathrm{d}  t}= -x_i + \sigma \bigg( \sum_{j=1}^{N}  w_{ij} x_j - \Theta_i \bigg).
\end{eqnarray}
In the above $x_i$ is the current state of $i$th neuron and  $w_{ij}$ is the connection strength, or {\it  weight}, between the neuron number $i$ and the neuron number $j$.  Each neuron is essentially  a threshold device with the threshold $\Theta_i$ or, in more general terms, a non-linear device, with the non-linearity described by the ``sigmoid" function $\sigma(z)$,  e.g. $\sigma(z)=\frac{1}{1+e^{-z}}$. This is one of the possible models for artificial NNs, and although it does not capture the real firing and spiking transmission processes observed in biological neurons, it  provides an approximate mathematical description of the most important ability of a NN -- the ability to recognize patterns, or to {\it classify}. 

The NN paradigm was a breakthrough in the field of Artificial Intelligence for the following reason. In conventional computing, two objects are regarded as the same only if they are identical. Therefore, to attribute a new pattern to an appropriate class (to recognize a pattern), a computer needs to know {\it all} elements that form the given class. This is not consistent with our everyday experience, in which living systems can successfully recognize patterns which they have never seen previously. This fundamental limitation was overcome by NNs as described below. 

If the function $\sigma$ and the thresholds $\Theta_i$ are fixed, 
the system (\ref{NN}) can be perceived as a non-linear dissipative dynamical system, whose vector field is determined by the weights $w_{ij}$. If the 
weights are symmetrical, i.e. $w_{ij}$$=$$w_{ji}$, one can introduce an energy function $E$ \cite{Hopfield_attractors_86}, such that the right-hand sides of Eq. (\ref{NN})
are the coordinates of the gradient of $E$. The function $E$ would typically have a number of local minima, each being a stable fixed point in the phase space with its own basin of attraction. 

{\bf Pattern recognition by a Hopfield NN with fixed weights.} Each minimum of energy $E$ represents the most typical or average representative of a certain class, or class centre. All patterns that belong to the same class are represented by the phase points in the basin of attraction of the respective stable fixed point. Since there are infinitely many points in the basin, there can be infinitely many patterns that belong to the same class, just like in reality. E.g. infinitely many projections of a certain flower,  registered by a cat looking at
it at different angles, are perceived as the same flower. 

An input pattern is represented by initial conditions in the phase space, which would fall in one of the basins of attraction available. Then the phase point  follows the vector field and moves towards the respective fixed point. When the fixed point is reached, the pattern is deemed recognized. 

{\bf Learning by a Hopfield NN.} Before the NN acquires the ability to classify, it needs to learn. Learning is understood as the adjustment of the weights $w_{ij}$, and in its turn the shaping of the energy landscape $E$. There exist a considerable number of algorithms to find the values of $w_{ij}$, see, e.g. \cite{Arbib} and references therein. Depending on the algorithm, learning in NNs can be supervised, semi-supervised \cite{Semi-Supervised}, reinforced  or unsupervised \cite{Hinton_connectionist_learning}. In any case, to train a NN, one presents it with a relatively large, but finite, number of example patterns. In {\it supervised} learning, the teacher also tells the NN how to classify each training pattern, i.e. manually attributes it to a certain basin of attraction. In addition, it specifies the total number of classes and the locations of the class centres, i.e. of fixed points. On the other extreme, in {\it unsupervised} learning, the NN is trying to figure out all fixed points and their basins on its own, by extracting some statistical information from the training set. {\it Unsupervised} learning presents the largest challenge out of all types of learning. 

Also, typically, a NN first learns and fixes its weights, and then performs recognition. However, there has been some effort in the direction of {\it on-line learning}, in which a NN would adjust its weights in the process of learning \cite{Online}.

{\bf Comparison with Hopfield NNs.} If continuous-time Hopfield NNs could learn in an unsupervised and on-line manner, they would work in the same way as the gradient self-shaping systems. 

{\bf Advantage over Hopfield NNs.} The existing algorithms used for the adjustment of weights in Hopfield NNs are quite good at developing the attractors (typically stable fixed points at the minima of the energy function) and of their basins of attraction,  in the right locations. However, 
whatever algorithm is used, it is very difficult, if not impossible, to control how the {\it whole} vector field changes in response to the training input. The largest problem is the occurrence of {\it spurious} minima, which develop by themselves as the weights are adjusted, and do not correspond to any valid classes. These minima affect pattern recognition, and this problem has still not been resolved after many years of effort. 

The desirable energy landscape should possess local minima at the points, where the most probable class representatives appear, and have no other minima. A function that would perfectly satisfy this condition is a PDD of all possible patterns, taken with a negative sign. And it is the PDD, that appears to be the energy in gradient self-shaping systems, albeit  smoothed by the kernel with a finite width. 
Thus, unlike Hopfield NNs, in the gradient self-shaping systems spurious minima do not occur. 

\subsection{Probabilistic neural networks}

The gradient self-shaping systems also have one feature in common with another type of NNs, called {\it probabilistic neural networks} \cite{Probabilistic_NN}. The purpose of the latter is to estimate the PDD of the incoming patterns, and then use it for classification purposes. Such NNs were developed in the attempt to overcome the spurious minima problem of the Hopfield NNs. 

The paradigm used here is essentially the same as in
all NNs: there is a collection of units with rigid architecture, and there are flexible/adjustable couplings between them. 
However, such NNs have a somewhat different architecture as compared to Hopfield NNs. Namely, in them there is always a separate layer of neurons,
such that each neuron codes a separate element of the training set. Thus, in order to take into account a new training pattern, one needs to physically add a new neuron to the system, thus making the whole system larger. In practice this implies that only a finite number of training patterns can be used, which imposes a considerable restriction on the system's performance. To lift the requirement of ``one pattern -- one neuron", this technique was improved \cite{Probabilistic_NN_smaller}, but the general idea remained the same: the system needs to be expanded to learn better. 

This paradigm in fact accounts for the popular ``grandmother neuron" hypothesis \cite{Grandmother_neuron}, which at the early ages of neuroscience suggested that in the brain the memory about a certain object was coded by a special neuron. 
E.g., the memory about one's grandmother has to be coded by the respective single neuron. This hypothesis contradicts the  Hopfield  NNs idea \cite{Hopfield_attractors_86}, that many memories can be coded by the same collection of neurons, as explained above. 

{\bf Comparison with probabilistic NNs.} Gradient self-shaping systems can do the same job as probabilistic NNs, i.e. to estimate the probability density distribution of incoming patterns and thus single out separate classes and their most typical representatives --  without supervision and on-line. 

{\bf Advantages over probabilistic NNs.} In estimating the PDD, the gradient self-shaping systems do not rely on the physical addition of new units in the course of learning, at least within the mathematical paradigm proposed. They can make use of as many training patterns as needed without any restrictions on their number. 

\section{Application to musical data.} 

\begin{figure}
\includegraphics[width=0.45\textwidth]{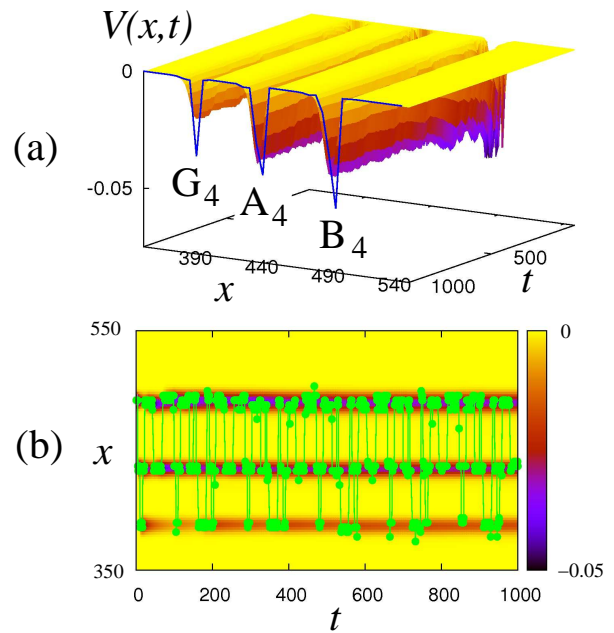}
\caption{\label{fig_flute_1d} (Color online.) Musical note recognition. (a) Evolution of the energy landscape $V(x,t)$ in response to a musical signal performed by an amateur musician. Local minima that develop eventually are very close to the frequencies of the musical notes $G_4$, $A_4$ and $B_4$ that enter the song.  (b)  Filled circles show the actual values of the input, and the shade of the background shows the depth of the energy function.  }
\end{figure}

Here, we illustrate how a gradient self-shaping system automatically discovers and memorises musical notes and phrases. 
A children's song ``Mary had a little lamb" was performed with a flute by an amateur musician six times.  The song involves three musical notes ($A$, $B$ and $G$), consists of 32 beats and was chosen for its simplicity to illustrate the principle. The signal was recorded as a 
wave-file  with sampling rate $8$kHz. In agreement with what is usually done in speech recognition \cite{Flanagan_77}, 
the short-time Fourier Transform was applied \cite{Allen_SFT} to the waveform with a sliding window of duration $\tau$$=$$0.75$ sec, which was roughly the duration of each note. The highest spectral peak was extracted for each window, which corresponded to the main frequency $f$ Hz of the given note. A sequence of frequencies $f(t)$ was used to stimulate the system (\ref{eq3}). 
Note, that each value of $f(t)$ was slightly different from the exact frequency of the respective note, because of the natural variability introduced by a human musician, and the signal $f(t)$ was in fact random, as seen from Fig. \ref{fig_flute_1d}(b).

Firstly, we illustrate how individual musical notes can be automatically identified. A one-dimensional system  (\ref{eq3}) received the signal $\eta(t)$$=$$f(t)$, resampled to $8$Hz to save computation time.  The function $f(t)$ can be seen as a realization of a 1st-order stationary and ergodic  process  $F(t)$, consisting of infinitely many repetitions of the same song, which we observe during finite time. This process has a one-dimensional PDD $p_1^F(f)$, which does not change in time.
A Gaussian kernel $g(z)$ was used with  $\sigma_z$$=$$\sqrt{5}$ Hz. As shown in 
Fig.~\ref{fig_flute_1d}(a), the energy converges to some PDD (with negative sign) shown by the solid line. It 
automatically discovers the most probable frequencies as follows,  figures in brackets showing the exact frequencies of the respective musical notes: 
434Hz (440Hz) for $A_4$, 490Hz (493.88Hz) for $B_4$, and 388Hz (392Hz) for $G_4$. 

Secondly, we show how the system (\ref{eq3}) can discover and memorize temporal {\it patterns} -- musical phrases consisting of four beats. The 4D ``foam" was used, and to each of its channels the same signal $f(t)$ was applied, but with a phase shift. Namely, at each time $t$ the system (\ref{eq3}) received a vector stimulus $\psi(t)$$=$$(f(t),f(t+\tau),f(t+2\tau),f(t+3\tau))$, $\tau$$=$$0.75$ sec. The procedure of creating a vector with the coordinates made of the delayed versions of the same signal is called delay embedding \cite{Takens_embedding}. For the purpose of this part, we can regard  $\psi(t)$ as a realization of a 4th-order stationary and ergodic vector random process $\Psi (t)$ (which we observe during finite time) with $4$-dimensional  PDD $p^{\Psi}_4(f_1,f_2,f_3,f_4)$. 
We used a multivariate Gaussian kernel $g$ with $\sigma_z$$=$$\sqrt{5}$ Hz in all of its four variables. 

\begin{figure}
\includegraphics[width=0.45\textwidth]{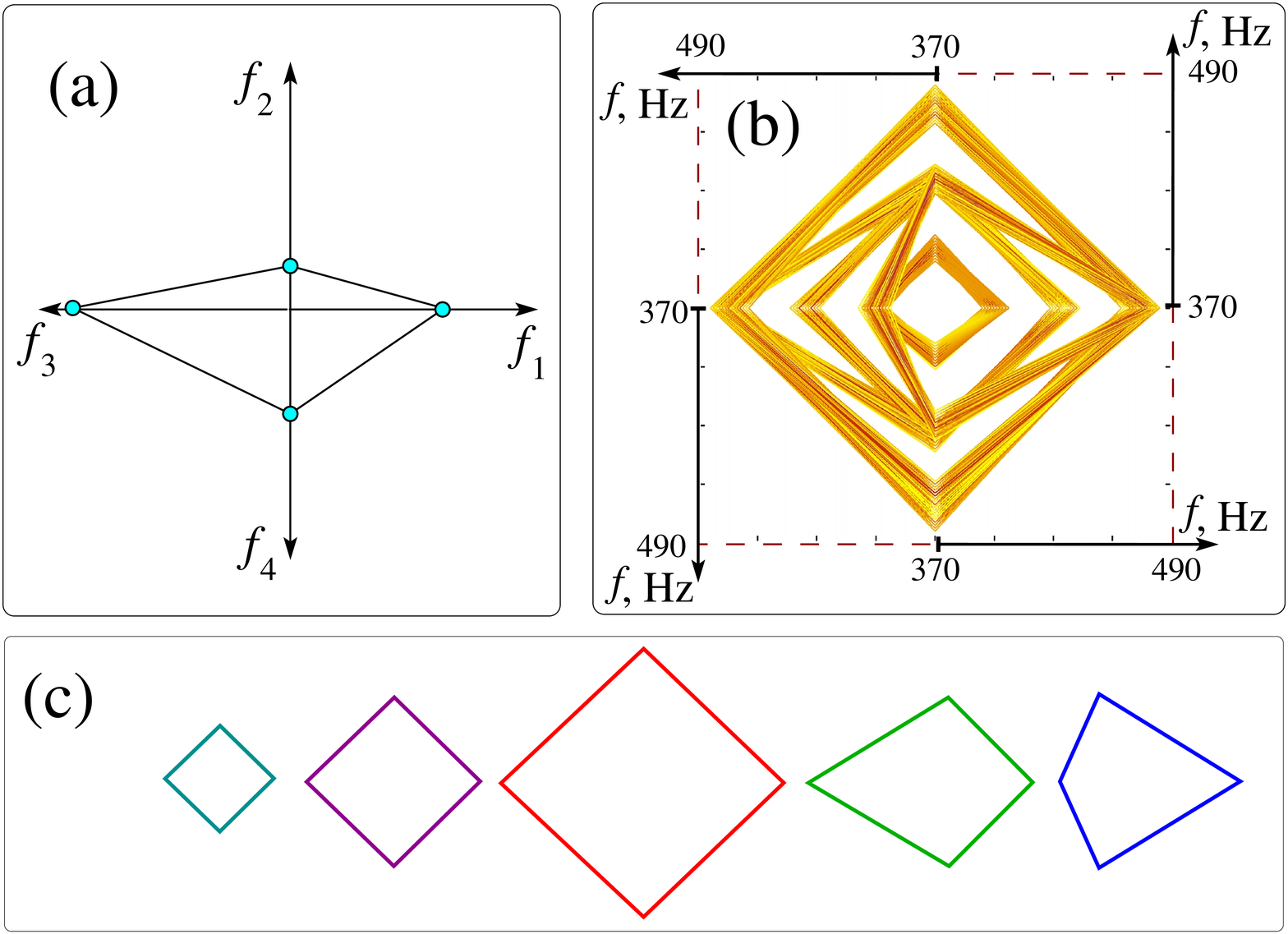}
\caption{\label{fig_flute_4d} (Color online.)  Musical phrase recognition. Description is in text.}
\end{figure}

One cannot visualize evolution of a 4D landscape in the same way as we did in Figs.~\ref{fig_uncor_cor}-\ref{fig_flute_1d}, and we
use an alternative representation. We take four half-axes and make their origins coincide (Fig.~\ref{fig_flute_4d}(a)). 
For each feasible input $\psi$$=$$(f_1,f_2,f_3,f_4)$ we put 4 points with coordinates $f_i$ on each of half-axes, and connect them by lines. Thus, any feasible input pattern is represented by a polygon on a plane. (This can be done for any dimension of input vector.) The value of $p^{\Psi}_4$ at each point can be represented by the color of the respective polygon (Fig.~\ref{fig_flute_4d}(b)). The polygon, whose color is the darkest, is the most probable pattern. Unfortunately, when too many polygons overlap, it might be difficult to see the darkest ones. But they can be found using the paradigm of a particle in the 4D landscape, that will go to one of the local minima representing one of  the most probable patterns: five such patterns are given in smaller scale in Fig.~\ref{fig_flute_4d}(c). Recognition of musical phrases is also illustrated by the supplementary  audio files \cite{URL}.

\section{Discussion and outlook} 

We started by proposing to treat information broadly as the shape of the matter, and the process of acquiring information, i.e. learning, as shaping of the matter in general. 
Staying within the dynamical systems framework, we  introduced a mathematical concept of a self-shaping dynamical system, which exploits these definitions of information and learning. We showed how such systems perform unsupervised learning and compare this mechanism with the one in the neural networks. 
The self-shaping systems  shape their velocity vector fields automatically under the influence of the external random stimulus. The resulting properties of the vector field, and consequently of the vector flow, are dictated by the statistical properties of the stimulus applied. We demonstrated how the simplest self-shaping systems of a gradient type develop the fixed point attractors together with their basins of attraction. We proved that for a stationary and ergodic input random process, the energy of such gradient systems converges to a smoothed probability density distribution of the input signal. The relevance of the new type of dynamical systems to the neural networks of two types is discussed. 
It is argued that the gradient self-shaping systems could serve the same purpose as neural networks, but would be lacking their limitations. 
The performance of a gradient self-shaping system is illustrated with an example in the form of a musical pattern. Namely, it is shown how
the system automatically discovers separate musical notes and musical phrases. 

Self-shaping systems of a gradient type, that were considered here, present only the simplest form of such systems. We predict that it will be possible to construct 
self-shaping systems that develop more complex attractors, such as limit cycles and chaotic attractors. Obviously, they would not be of a gradient type. Finding the general mechanisms of their formation will be the subject of our future work. 

What we present here is a mathematical proposal for the systems of a new class. We argue that, if implemented in hardware, such systems would have considerable advantages over neural networks. However, the physical principles upon which such systems could be built are not obvious at the moment. Therefore, this proposal represents an engineering challenge  and calls for the development of the devices of a new kind.

{\bf Self-organization and self-shaping.} A very important property of non-linear systems, both natural and man-made, is their ability to self-organize. Some famous examples are Benard cells \cite{Benard_cells} that automatically form in a heated liquid, and Belousov-Zhabotinsky chemical reaction \cite{Belousov_reaction}, in which the liquid spontaneously changes colour. 
In terms of dynamical systems, self-organization has been traditionally understood as automatic shaping of the {\it solutions} that start from a range of initial conditions, given the fixed structure of the vector field and/or of its perturbations.  We now wish to extend the self-organization principle to the automatic shaping of the vector field itself.  
It most vividly manifests itself in  living systems, that continuously change themselves in response to external influence. Therefore,
the suggested self-shaping approach might prove a helpful paradigm when modelling adaptation and development in living systems in general. 

\begin{acknowledgments}
The authors are grateful to Alexander~Balanov for thorough reading and helpful critical comments on all drafts of this paper, 
to Mark Robbins, Victoria Marsh and Scott Dickson for their feedback on the paper, 
and to Victoria~Marsh for playing the flute. 
\end{acknowledgments}

\end{document}